**Title of the article:**

Governance of Ethical and Trustworthy AI Systems: Research Gaps in the ECCOLA Method

**Authors:**

Mamia Agbese, Hanna-Kaisa Alanen, Jani Antikainen, Erika Halme, Hannakaisa Isomäki, Marianna Jantunen, Kai-Kristian Kemell, Rebekah Rousi, Heidi Vainio-Pekka, Ville Vakkuri

**Notes:**

- This is the authors' version of the work
- The definite version was published in: Agbese, M., Alanen, H.-K., Antikainen, J., Halme, E., Isomäki, H., Jantunen, M., Kemell, K.-K., Rousi, R., Vainio-Pekka, H., & Vakkuri, V. (2021). Governance of Ethical and Trustworthy Al Systems : Research Gaps in the ECCOLA Method. In T. Yue, & M. Mirakhorli (Eds.), REW 2021 : 29th IEEE International Requirements Engineering Conference Workshops (pp. 224-229). IEEE. https://doi.org/10.1109/REW53955.2021.00042

- Copyright owner's version can be accessed at https://doi.org/10.1109/REW53955.2021.00042

# Governance of Ethical and Trustworthy AI Systems: Research Gaps in the ECCOLA Method

Mamia Agbese
Faculty of Information Technology
University of Jyväskylä
Jyväskylä, Finland
mamia.o.agbese@jyu.fi

Hanna-Kaisa Alanen
Faculty of Information Technology
University of Jyväskylä
Jyväskylä, Finland
0000-0002-8797-3432

Jani Antikainen
Faculty of Information Technology
University of Jyväskylä
Jyväskylä, Finland
0000-0003-3367-0492

Erika Halme
Faculty of Information Technology
University of Jyväskylä
Jyväskylä, Finland
0000-0003-0750-1580

Hannakaisa Isomäki
Faculty of Information Technology
University of Jyväskylä
Jyväskylä, Finland
0000-0002-6021-3118

Marianna Jantunen
Faculty of Information Technology
University of Jyväskylä
Jyväskylä, Finland
marianna.s.p.jantunen@jyu.fi

Kai-Kristian Kemell
Faculty of Information Technology
University of Jyväskylä
Jyväskylä, Finland
0000-0002-0225-4560

Rebekah Rousi
Faculty of Information Technology
University of Jyväskylä
Jyväskylä, Finland
0000-0001-5771-3528

Heidi Vainio-Pekka
Faculty of Information Technology
University of Jyväskylä
Jyväskylä, Finland
0000-0002-9736-3400

Ville Vakkuri
Faculty of Information Technology
University of Jyväskylä
Jyväskylä, Finland
0000-0002-1550-1110

*Abstract*— **Advances in machine learning (ML) technologies have greatly improved Artificial Intelligence (AI) systems. As a result, AI systems have become ubiquitous, with their application prevalent in virtually all sectors. However, AI systems have prompted ethical concerns, especially as their usage crosses boundaries in sensitive areas such as healthcare, transportation, and security. As a result, users are calling for better AI governance practices in ethical AI systems. Therefore, AI development methods are encouraged to foster these practices. This research analyzes the ECCOLA method for developing ethical and trustworthy AI systems to determine if it enables AI governance in development processes through ethical practices. The results demonstrate that while ECCOLA fully facilitates AI governance in corporate governance practices in all its processes, some of its practices do not fully foster data governance and information governance practices. This indicates that the method can be further improved.**

*Keywords—AI, Ethics, Ethical AI, ECCOLA, AI governance, ML*

I. INTRODUCTION

Artificial Intelligence (AI) is arguably one of the promising technologies of the current decade. Integration with machine learning (ML) has further revolutionized AI technology to improve the functionality of AI systems [1]. Consequently, AI systems are increasingly employed in various sectors, with their proliferation in critical application areas such as medicine, transportation, and security, raising ethical concerns [2]. The ethics of AI deals with the moral behavior of humans in the design, usage, and behavior of machines [10]. Many ethical issues have been identified with the use of AI systems [13]. Some of which include AI systems usage can lead to job loss for humans, propagate bias, invade privacy, undermine fairness practices, thwart accountability, or be misused by malicious actors to perpetuate evil [3]. These issues represent some of the ethical concerns that can impede AI progress [4].

Researchers, governments, and organizations in the ethical community continue to make progress in mitigating these ethical concerns by producing guidelines and frameworks for developing ethical AI systems that users can trust [1]. Some of these guidelines include the High-level Expert Group on Artificial Intelligence (AI HLEG), the Expert Group on AI in the society of the Organization for Economic Co-operation and Development (OECD), the Initiative for Ethically Aligned Design (EAD) for autonomous and intelligent systems by (IEEE) and the Advisory Council on the Ethical Use of Artificial Intelligence and Data in Singapore [1]. However, developers still struggle to effectively transition these guidelines to trustworthy AI systems [5]. According to [6], one of the weaknesses of the principles and guidelines approach to Trustworthy ethical AI systems is a lack of proven methods that translate principles to practice. Furthermore, existing method guidelines for implementing trustworthy AI systems are considered challenging due to their lack of practicality in implementation and inability to translate principles to design [6].

Another area of concern gaining traction in the development of ethical AI that may pose a barrier to the progress and adoption of AI is the regulation and governance of AI systems [2]. According to [7], ethical principles alone are insufficient for developing and deploying ethical and trustworthy AI. AI systems require strong governance controls that manage processes and create associated audits that enforce principles [7]. "Reference [28] explain" that issues such as scalability can pose an ethical issue in the governance of AI systems [28]. AI systems exhibit characteristics peculiar to their structure and architecture, posing an ethical challenge in their governance [28]. Scaling AI systems using technology such as 5G can lead to challenges on the placement of logic that will govern the system [28]. In addition, AI systems must consider several factors and actors to reconcile conflicting forces [28], which can challenge its governance.

Currently, the topic of AI governance is widely unexplored [9], with varying studies on AI governance and regulatory issues [4, [6, 10, 12]. These may be attributed to AI systems being a global issue and not a one-size-fits-all recommendation with varying practices regarding context and culture [4]. Ethics on its own is subject to cultural interpretation, making it unstable ground for generic regulation [11]. Therefore, AI developers and designers face the challenge of ensuring AI systems are ethical and facilitate governance processes that enable effective audits [7]. These types of practices can help improve the widespread adoption of AI systems and not impede their progress [4]. Hence a need to examine ethical AI development methods to determine their facilitation of AI governance.

In this context, we explore ECCOLA, a development method for Trustworthy AI systems [5] to ascertain its capacity to account for and facilitate AI governance. This paper aims to analyze the method to identify if governance practices exist and are fostered by ECCOLA and expose gaps where they are lacking. The analysis will also assess ECCOLA's ability to promote AI governance through its ethical practices. The information collated can help to improve the method further and provide insight into the governance of trustworthy AI development methods.

The rest of the paper is structured as follows. First, we present the theoretical background in section II. Section III outlines the analysis, which includes the utilized methodology, findings, and limitations. Finally, section IV presents the conclusion and further research.

## II. THEORETICAL BACKGROUND

### A. AI Ethics

AI ethics is described by [7] as the practice of using AI with good intentions in empowering employees and businesses with a positive impact on customers and society. Most of the concerns raised in the previous section represent a societal stand on AI challenges. Therefore, ethics helps in engendering trust and scaling AI technology with confidence [7]. According to [24], ethical AI is not intended to give machines responsibilities for their actions and decisions. Instead, it gives people and organizations more responsibility and accountability for their actions and decisions [24]. There is an increased demand for more ethical and explainable AI systems [23, 24]. Humans are becoming more reticent in adopting technologies they cannot directly interpret, explain, and audit. But, developers believe that more transparent or explainable AI models are less accurate [4]. They imply a trade-off between accuracy and explainability [4]. However, this is yet to be proven [4].

### B. ECCOLA Method

ECCOLA method is an example of transitioning principles to practice [5]. The method has been developed [8] to bridge the gap between ethical research and practice and incorporate ethical guidelines from the IEEE and the EU Trustworthy AI guidelines [5]. The ethical guidelines in ECCOLA represent the main pillars of ethics [6]. They also form a fragile consensus of a shared foundation to build ethical AI systems [6]. ECCOLA method comprises 21 cards split into eight themes [5]. The themes are *Analyze*, *Transparency*, *Data*, *Agency and Oversight*, *Safety and Security*, *Fairness*, *Wellbeing*, and A*ccountability [5]*. Each theme consists of one to six cards, with each card providing a more detailed approach to the theme it represents [5]. ECCOLA aims to create trustworthy AI systems by providing developers with an actionable tool for implementing AI ethics [5]. The method employs a human-centered approach that requires human actors to be the clear focus [5]. The methodologies are designed to reflect this [5]. For ethical concerns, ECCOLA offers insight for solutions [5]. It provides a practical approach that can be easily integrated into any existing method by asking questions that enable developers to consider the various ethical issues present in developing ethical AI systems [5]. For non-functional use, ECCOLA aids in the creation of user stories for product owners in ethical matters and facilitates communication for developers [5].

ECCOLA is part of the software development process and falls under development methods and practices for implementing ethically aligned AI systems [5]. It uses developed principles and concepts [5] to provide a practical tool for software developers, product managers, and consultants in their EAD for AI systems [5]. ECCOLA aims to go beyond the misconception that operational tools alone are sufficient for development as a software method tool [26]. It aligns with successful method development and deployment components and offers insight into the change management involved in using the method [26]. It asks software developers questions from the initial stage (conception), the mental stage (where thinking tools, principles, and patterns are analyzed), to the operational stage, which includes the life-cycle of the AI systems as seen in Fig. 1 [26].

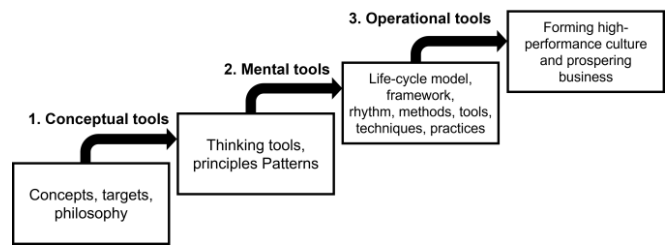

Fig.1. Necessary components for method adoption, development, and deployment [26]

ECCOLA is a novel method and has not been subject to further analysis beyond its development and validation. Therefore, this study seeks to analyze ECCOLA to improve its robustness, enabling its widespread adoption [16]. For method models to attain robustness, a need exists for evaluations to benchmark areas that have not been taken into consideration or fully exploited [16, 25]. In addition, a lack of robustness in a method can lead to duplicated efforts with little practical benefits slowing the pace of research [25]. Therefore, it is pertinent to evaluate ECCOLA to ascertain what governance practices it facilitates and lacks. This information can help refine the method and improve its effectiveness, functionality, and ultimately its robustness [17].

*C. AI Governance*

According to [12], "AI governance is about AI being explainable, transparent, and ethical" [p. 1]. He explains that AI governance or the governance of AI requires accountability from all liable actors involved in developing, deploying, and using AI systems [12]. These actors' clear and transparent accountability measures can facilitate effective tracking and measurement, aiding AI governance practices [12]. However, some of these guiding principles are subjective and may result in different interpretations of AI governance [11]. For example, transparency may refer to software codes or algorithms in technical terms, translating to an entirely different concept in policy and regulations [11]. These inconsistencies often result in tensions in AI governance approaches, with one of these crucial elements left out or lacking in specificity [11].

Nonetheless, AI governance helps to address ethical, security, and regulatory issues in AI systems [2]. A mishap could have monumental consequences for sensitive sectors like healthcare, where individuals' privacy and security of personal health information is paramount [2]. Appropriate regulatory and accreditation measures in governance practices incorporated in AI systems can help safeguard some of these threats [2]. Therefore, governance models grounded on ethical components such as fairness, transparency, trustworthiness, and accountability are advocated for AI systems [2].

The governance of AI involves the study of how humanity navigates AI transitions within all sectors, including political, ethical, economic, and military [13], [14]. This transition is essential as it provides interaction of AI governance with all these sectors in order to make it effective [14]. Transition entails that AI governance is broken down into different components, such as the technical component, which deals with the technical aspects of AI systems like data, information, and information asset security [13],[14]. Other components can include the political component that deals with regulation, standardization, and political dynamics between actors involved in developing AI systems [13]. This political dynamic can include corporate governance and ideal governance for governing transitions to advanced AI, which is out of the scope of this research [13].

To help improve AI governance, AI development methods should ensure transparent, accountable, and explainable practices [12], which may require governance mechanisms that minimize risks and probable downsides to be incorporated to harness the full potential of AI systems [12] [15]. According to [13], AI systems should be regarded as ethical impact agents in AI governance and subjected to an ethical risk assessment to reduce the impact of any ethical exposure. However, including a governance structure for developers can be challenging as AI systems constantly evolve and regulations across the board differ [12, 15]. Therefore, it is essential to take a flexible and layered approach that accommodates different governance practices as AI technology grows and develops [12], 17].

### III. ANALYSIS

In this section, we present an analysis of ECCOLA through the lens of AI governance to determine if the governance practices in the method facilitate AI governance. As discussed earlier in section I, AI governance is still largely unexplored, with a diverse discourse on a consented framework that AI developers can adopt [13]. These discussions focus on a consented framework and others on consented regulations to aid AI governance [13]. Consequently, several studies have developed different AI governance models to suit the particular sector the research focuses on [2, 9, 13, 24]. There is a lack of consensus on AI governance frameworks for developers at the higher level, where ECCOLA typically operates. As a result, a typology approach by [10] is employed in the analysis of ECCOLA.

*A. Methodology*

The typology approach aims to design a typology that will match identified AI governance practices in ECCOLA to its corresponding ethical theme [10]. According to [10], this approach presents a pragmatic view. This view involves a rationalization process of what is regarded as fair consideration based on the discussion on the analysis results [10]. Thus, the methodology in the study is taking a pragmatist stance towards evaluating the ECCOLA model using the typology approach. The identified AI governance practices will serve as a medium to identify, check and examine the cards in ECCOLA to provide a fair consideration [10]. This approach can help detect the governance practices that exist in each of the 21 ECCOLA cards. Creating the typology requires that the ethical themes in ECCOLA are combined with AI governance practices [10]. To help identify suitable AI governance practices, we used the AI Governance Framework by The Personal Data Protection Commission in Singapore (PDPC) [18, 10].

The AI Governance Framework employs a human-centric ethical approach towards AI governance practices to create an AI ecosystem jointly and inclusively for actors dealing with AI [18]. The Framework has been validated and employed by international organizations such as Facebook [18]. In addition, it has received feedback from the European Commission High-Level Expert Group and the OECD expert group on AI [18]. The Framework focuses on four broad governance areas: Internal governance structures and measures, Human involvement in AI-augmented decision making, Operations management, and Stakeholder interaction and communication [18]. *Internal governance decision-making operations* represent internal governance measures and structures that integrate risks, values, and responsibilities relating to AI decision-making [18]. *Determining human involvement in AI-augmented decision-making* deals with methodologies to help determine acceptable risk appetite for AI systems and the level of human participation in AI-augmented practices or decisions [18]. *Operations management* deals with matters considered in developing, selecting, and maintaining AI systems, including data management, Information management, and corporate management. [18]. *Stakeholder interaction and communication* deals with communication strategies with stakeholders and effective management [18]. These broad AI governance areas comprise different governance practices, measures, and frameworks that navigate the course to effective AI governance [18]. Each key area comprises guidance on measures that promote responsible governance practices for the use of AI that can be adopted or set as applicable [18]. For example, suppose a governance goal is defined. In that case, clear guidance appropriate to the realization of these goals is applied to roles and responsibilities for all parties and activities to achieve the set goals [18]. This guidance can also provide accountability for tracking progress [18]. The AI model Framework provides a flexible approach and can be adapted to suit corresponding needs considering the relevant elements [18].

We leverage the flexibility of the AI governance framework for the ECCOLA analysis. The four model practices of the AI governance framework are streamlined into three governance practices identified in the four broad governance areas. The practices are corporate governance, information governance (IG), and data governance (DG). This selection is based on how invaluable data, information, and corporate governance activities are to developing AI systems and the need for principles or guidance to assist with their governance [30]. In addition, the practices of IG, corporate governance, and data governance are crucial for developing AI systems due to data sensitivity (the need for fair and explicable data) [28], the relevance of trustworthy information, and the need for an overarching level of corporate supervision in their management and governance [30]. Also, some of the governance guidance in the AI model framework corresponds with the governance principles of these three governance practices identified in the literature review [18], [19], [20]. Overall, these practices are essential for developmental methods like ECCOLA [13]. According to [29], the governance of AI can be interpreted differently based on perspective. They explain that technical researchers can consider it from a technical viewpoint, while regulators can view it from the perspective of trustworthiness [28]. As such, it can be broken down into manageable structures that enable governance [18]. Therefore our selection is explained as AI governance not being considered a mere process but a set of important aspects that need to be considered in development methods [28]. Hence this study views AI governance from the viewpoint of Corporate governance, IG, and DG [28].

The result of combining the ethical themes in ECCOLA with AI governance practices is shown in (Table 1), which will serve as the analytical template for examining the ECCOLA cards. Each card will be examined in line with the ethical theme in ECCOLA and the governance practices. AI governance is represented horizontally and the ethical themes in ECCOLA vertically.

Table 1. Ethical Theme In ECCOLA And AI Governance Practices [6]

| Ethical themes in ECCOLA | Governance Practices | | |
|---|---|---|---|
| | Data Governance | Information Governance | Corporate Governance |
| Analyze | | | |
| Transparency | | | |
| Data | | | |
| Agency and Oversight | | | |
| Safety and Security | | | |
| Fairness | | | |
| Wellbeing | | | |
| Accountability | | | |

A literature review was carried out to identify governance guidelines for AI. The keywords include "AI Information governance practices," "AI Information governance principles," "AI data governance practices," "AI data governance principles," "AI corporate governance practices," "AI corporate governance principles" searched with "ethical artificial intelligence" OR "ethical AI development (design) methods." This process would help establish clear guidance measures for the analysis, as recommended by [18]. The review revealed emerging efforts in standardizing data and IG practices [30] but with no clear consensus. Some of the studies [22], [23], [30] reveal the need for IG and data governance frameworks, but no clear governance principles were identified to facilitate the governance process. Therefore to help identify principles that enable governance practices, a broader review was employed. The keywords used in the review include "data governance," "information governance," "information governance practices," "data governance practices," "corporate governance," "corporate governance practices". All searched with "ethical AI" OR "ethical AI development (design) methods." The findings revealed a common consensus for data and information governance practices. The popular, published guidance includes the Principles® or Generally Accepted Recordkeeping Principles® (GARP®) by ARMA International [19] and The DAMA Guide to the Data Management Body of Knowledge (DAMA-DMBOK) by the Data management association (DAMA) [20]. These principles are globally accepted, flexible, easily incorporated with their practices employed and adapted across virtually all sectors and industries [19], [20]. No suitable general framework was identified; therefore, a list was compiled from various corporate governance guidelines.

The second task involved identifying tools and methods to fill out the typology [10]. According to [10], there are many ways to fill out the typology, which may involve already established tools and methods, interviews, surveys, or literature reviews. However, this study is based on ECCOLA, method; therefore, each of the 21 cards' practices will be used to fill out the typology. The final result is presented in (Table 2). The final task was to evaluate the ethical practices in each ECCOLA card with the governance guidance outlined from the reviewed sources to create the typology. This evaluation was achieved using conceptual content analysis.

The first step was to identify a governance practice in the card and then match the corresponding card to fit in the typology [10]. This process requires each high-level ethical principle to be translated for clarity [10]. However, in ECCOLA, the high-level ethical principles have already been translated into eight ethical themes. Therefore, for further clarity, we matched the themes in each card to the ethical theme from which it originated [10]. Creating the typology involved using content analysis to analyze the governance practices in GARP, DAMA, and corporate governance principles to determine a sample for analysis [31]. Then each ECCOLA card was coded into a manageable content category and analyzed to identify the presence of governance words and concepts similar or matching the GARP, DAMA, and corporate governance principles. The cards were coded as existing or not based on the governance practices described in the cards that indicate or lead to governance practices in line with the identified principles [31]. Each card identified with practices corresponding to a particular governance principle was matched. The final analysis (Table 2) aims to synthesize the governance practices that exist in the ECCOLA method [10].

*B. Discussion*

*Identification of AI governance practices:* The study reveals that all three governance practices are present within the activities in the ECCOLA cards. Furthermore, each card promotes one or more governance practices in the recommended actions. For example, one of the guidance from the DAMA governing practices stipulates the essence of data quality by recommending practices such as defining, monitoring, maintaining data integrity, and improving data quality [21]. ECCOLA card #8 Data Quality asks questions like What are good or poor quality data in your system? [5], How is the quality and integrity of data evaluated? [5]. Asking these questions aids developers ensure they adhere to governance practices. According to [22], AI governance practices that effectively manage and govern data and information are the foundation of trustworthy AI. In addition, DG practices in the development of AI systems can lead to the generation of information assets that reflect IG practices, facilitating effective auditing for corporate governance practices [22].

*Distribution of AI Governance practices:* Among the three, corporate governance emerged as the dominant practice, data governance in second place, and IG. All the cards in ECCOLA exhibited corporate governance practices. Indicating accountability practices; according to [23], ethics and corporate governance practices are vital for developing sustainable AI that users can trust. Therefore, incorporating these practices in development methods can help ensure that AI systems are geared towards sustainability, trustworthiness, and AI systems that foster audit [23]. On the other hand, data and IG practices exist but not in the same proportion as corporate governance. Therefore, a need may exist for a review of ECCOLA to reflect these practices.

*Ethical themes and AI governance practices:* The evaluation indicates that the Transparency theme accounts for most of the governance practices in ECCOLA. The theme identified the most cards in all three governance practices. Transparency deals with open and explainable measures in the development of ethical AI systems [23]. Transparency in development methods enables accountability measures that aid audits and foster good AI governance [23]. According to ]23], scaling autonomous AI systems can be complex due to their uncertain and dynamic learning nature. Therefore, transparent mechanisms in their development can aid monitoring, auditing, and subsequent governance to ensure that the systems remain trustworthy [23]. Furthermore, accountability fosters audit practices essential in the governance of AI systems [23], [25].

*C. Results*

The main aim of this study is to analyze ECCOLA to identify gaps that can lead to areas of improvement. The principal research gap identified is related to IG practices. The analysis revealed that IG has the least representation in ECCOLA. The identified governance practices exist in ECCOLA but to varying degrees. Corporate governance practices are represented in all the cards, but data governance and IG practices are not completely represented in all the cards. This indicates that ECCOLA can be improved. Therefore, ECCOLA can be further analyzed in-depth using an appropriate framework to integrate IG practices fully to improve its robustness. The result also indicates that ECCOLA facilitates AI governance through its ethical practices.

Table 2. Analysis of ECCOLA and AI governance practices [6]

| Ethical themes in ECCOLA matched with corresponding themes in cards | Governance practices | | |
|---|---|---|---|
| | Information Governance | Data Governance | Corporate Governance |
| **Analyze** | | | |
| Analysis of Stakeholder participation | Card #0 | Card #0 | Card #0 |
| | | | |
| **Transparency:** | | | |
| Types of Transparency | Card #1 | Card #1 | Card #1 |
| Explainability | Card #2 | Card #2 | Card #2 |
| Traceability | Card #5 | Card #5 | Card #5 |
| Communication | | | Card #3 |
| Documenting trade-offs | Card #4 | Card #4 | Card #4 |
| Systems Reliability | Card #6 | | Card #6 |
| | | | |
| **Data** | | | |
| Privacy and Data | | Card #7 | Card #7 |
| Quality of Data | | Card #8 | Card #8 |
| Access to data | Card #9 | Card #9 | Card #9 |
| | | | |
| **Agency and Oversight;** | | | |
| Human Agency | | | Card #10 |
| Human oversight | | | Card #11 |
| | | | |
| **Safety and Security:** | | | |
| System security | Card #12 | Card #12 | Card #12 |
| | | | Card #6 |
| System Safety | | | Card #13 |
| | | | |
| **Fairness**: | | | |
| Accessibility | | | Card #14 |
| Stakeholder participation | Card #15 | | Card #15 |
| | | | |
| **Wellbeing:** | | | |
| Social Impact | | | Card #17 |
| Environmental Impact | | Card #16 | Card #16 |
| | | | |
| **Accountability**: | | | |
| Auditability | Card #18 | Card #18 | Card #18 |
| | | | Card #4 |
| Ability to Redress | | | Card #19 |
| Minimization of negative Impacts | Card #20 | Card #20 | Card #20 |

*D. Limitation*

This study explored AI governance based on general practices of data, information, and corporate governance. However, it is important to note that AI governance encompasses various practices that can be further explored. In addition, the evaluation was carried out mainly to identify AI governance practices in ECCOLA and not to evaluate them in-depth. Therefore, the interpretation was subject to the context of ECCOLA and the practices in the cards. As such, an in-depth approach to the identified practices can provide a different insight into the results.

Another potential limitation is related to the content validity of the AI governance practices. In this study, the governance practices are based on existing governance practices in literature adapted to suit AI governance since there is no consensus on a general AI governance framework. However, the governance practices are flexible and were taken into consideration accordingly.

IV. CONCLUSION AND FUTURE WORK

Continuous progress and development in ML technology will likely translate to increased scaling of AI systems as socio-technical and cyber-physical systems [21]. While the ethics community continues to navigate the development of ethical AI systems to help stem various ethical concerns, it is essential that the emerging and burgeoning governance concerns are equally addressed. AI governance has proven challenging [21] due to the broad scope of the different technologies and layers it encompasses. However, the need for its application is invaluable primarily as developers continue to scale and produce autonomous and semi-autonomous AI systems [22].

Therefore, the evaluation of ECCOLA concerning AI governance is essential for the method's robustness [26]. The practices of data governance, IG, and corporate governance identified in ECCOLA suggest that ethical principles are essential for governance practices [23]. The Transparency theme plays an essential role in AI governance in ECCOLA, suggesting that transparent practices engender effective audits that can lead to AI governance [23]; this can translate to other ethical development methods and software development methods for AI [27, 28, 29,]. The research further reveals an uneven distribution of these practices, with corporate governance emerging as the dominant practice, indicating that all the ECCOLA cards can be further improved to completely incorporate data and IG practices in all the cards.

Future research should consider an in-depth analysis of IG, the least represented practice, with an appropriate framework such as the GARP. Such an in-depth analysis can reveal which of the cards need to be improved. Research can also be carried out to identify the relationship of the other ethical principles with AI governance and why they are not as dominant as the principles of transparency.